\documentclass[aps,prl,
              reprint,
              amsmath,amssymb,amsfonts,groupedaddress,
              twocolumn,
              showpacs,
              superscriptaddress,floatfix]{revtex4-1}

\usepackage{graphicx}
\usepackage{dcolumn}
\usepackage{bm}

\usepackage{color}

\begin{document}

\title{Plasma-Based Generation and Control of a Single Few-Cycle\\
High-Energy Ultrahigh-Intensity Laser Pulse}

\author{M. Tamburini}\email{matteo.tamburini@mpi-hd.mpg.de}
\affiliation{Max-Planck-Institut f\"ur Kernphysik, Saupfercheckweg 1, D-69117 Heidelberg, Germany}
\author{A. Di Piazza}
\affiliation{Max-Planck-Institut f\"ur Kernphysik, Saupfercheckweg 1, D-69117 Heidelberg, Germany}
\author{T. V. Liseykina}
\affiliation{Institut f\"ur Physik, Universit\"at Rostock, D-18051 Rostock, Germany}
\author{C. H. Keitel}
\affiliation{Max-Planck-Institut f\"ur Kernphysik, Saupfercheckweg 1, D-69117 Heidelberg, Germany}

\date{\today}

\begin{abstract}
A laser-boosted relativistic solid-density paraboloidal foil 
is known to efficiently reflect and focus a counterpropagating laser pulse. 
Here we show that in the case of an ultrarelativistic counterpropagating pulse, 
a high-energy and ultrahigh intensity reflected pulse can be more effectively 
generated by a relatively slow and heavy foil than by a fast and light one. 
This counterintuitive result is explained with the larger reflectivity 
of a heavy foil, which compensates for its lower relativistic Doppler factor. 
Moreover, since the counterpropagating pulse is ultrarelativistic, 
the foil is abruptly dispersed and only the first few cycles of the 
counterpropagating pulse are reflected. Our multi-dimensional particle-in-cell simulations 
show that even few-cycle counterpropagating laser pulses can be further shortened 
(both temporally and in the number of laser cycles) with pulse amplification. 
A single few-cycle, multi-petawatt laser pulse with several joule of energy 
and with peak intensity exceeding $10^{23}\text{ W/cm$^2$}$ can be generated 
already employing next-generation high-power laser systems. 
In addition, the carrier-envelope phase of the generated few-cycle pulse 
can be tuned provided that the carrier-envelope phase of the initial 
counterpropagating pulse is controlled.
\end{abstract}

\pacs{52.59.Ye, 52.38.-r, 52.65.Rr, 42.65.Re}

\maketitle

A wide range of novel studies in nonlinear optics 
as well as the major new regimes of extreme field physics 
require laser pulses which simultaneously exhibit the following three key features: 
few-cycle duration, high-energy and ultrahigh intensity. 
Already in nonrelativistic atomic physics, 
it has been demonstrated that quantum processes can be controlled 
by manipulating the pulse shape of few-cycle laser pulses~\cite{krauszRMP09}. 
In order to achieve the same goal also in the ultrarelativistic regime 
and in the realm of nonlinear QED, few-cycle laser pulses with 
tunable carrier-envelope phase (CEP) are required with peak intensities 
largely exceeding $10^{20}\text{ W/cm$^2$}$~\cite{meurenPRL11,titovPRL12,dipiazzaRMP12}. 
At such high intensities, for example, the nonlinear Compton emission spectrum 
is expected to show pronounced pulse-shape effects~\cite{mackenrothPRA11,bocaPRA12}. 

Although next-generation 10-PW optical laser systems 
are expected to generate laser pulses with 150-300~J energy 
and 15-30~fs duration~\cite{korzhimanovPU11,dipiazzaRMP12} 
[full-width-at-half-maximum (FWHM) of the pulse intensity], 
the limited bandwidth renders the generation of few-cycle pulses 
with multi-joule energy very challenging~\cite{herrmannOL09,witteIEEE12}. 
Indeed, the only laser system aiming at 1-PW power and few-cycle 
duration is the Petawatt Field Synthesizer~\cite{pfsURL}. 
Several methods for further shortening and amplifying 
laser pulses have been proposed, e.g., Raman~\cite{malkinPRL99,torokerPRL12} 
and Brillouin~\cite{lanciaPRL10,weberPRL13} backscattering, 
interaction with plasma waves~\cite{faurePRL05,schreiberPRL10} 
and ionization induced self-compression effects~\cite{wagnerPRL04,skobelevPRL12}. 
However, none of the pulses generated employing the above-mentioned methods 
simultaneously exhibit few-cycle duration, multi-joule energy 
and ultrarelativistic intensity. 
In fact, the initial intensity is bounded to relatively moderate values 
and the generated pulses are transversely and temporally modulated, which 
might prevent their subsequent focusing to ultrarelativistic intensities. 
In addition, the CEP control, which is crucial for many applications, 
has not been demonstrated in any of the above-mentioned methods. 

In this Letter, we put forward the concept of a laser-boosted solid-density 
paraboloidal relativistic ``mirror'', interacting with a superintense counterpropagating 
laser pulse, to generate a CEP tunable few-cycle pulse 
with multi-joule energy and peak intensity exceeding $10^{23}\text{ W/cm$^2$}$. 
Contrary to intuition, it is found that a heavy and therefore 
relatively slow ``mirror'' should be employed to maximize the intensity 
and the energy of the reflected pulse, since its larger reflectivity 
compensates for the lower velocity. 
Furthermore, the short duration of the reflected pulse 
is achieved by employing a superintense incident pulse, 
which abruptly disperses the plasma mirror after only the first few cycles. 
Multi-dimensional particle-in-cell (PIC) simulations indicate 
both the feasibility of the presented setup by employing next-generation 
multi-PW laser systems and a considerable shortening 
with amplification even for already few-cycle laser pulses. 

In the proposed setup, a ``driver'' pulse with frequency $\omega$ and 
(average) intensity $I_d$ accelerates a ``mirror'' to 
relativistic velocities along the positive $x$ direction 
and a ``reflected'' pulse is generated in the collision of the mirror 
with a counterpropagating ``source'' pulse, also with frequency $\omega$ 
and with intensity $I_s$. Here and below, the subscript $s$ ($d$) and the upper (lower) 
sign refer to the source (driver) pulse counterpropagating 
(copropagating) with respect to the mirror, and $T=2\pi/\omega$ ($\lambda=cT$) 
is the laser period (wavelength). Our aim here is to 
determine the conditions for maximizing both the intensity 
and the energy of the reflected pulse. In order to develop 
an analytical model, for the thin foil we employ the Dirac-$\delta$ density profile 
$n(x)=n_e\ell\delta(x)$~\cite{vshivkovPP98,macchiNJP10}, 
where $n_e$ and $\ell$ are the foil density and thickness, respectively.
If the foil moves with velocity $v_x=\beta c > 0$, its reflectivity is given by 
$\mathcal{R}_{s/d} = \zeta_{s/d}^{2}/(\zeta_{s/d}^{2} + \Gamma^2_{s/d})$~\cite{macchiNJP10}, where 
$\Gamma^2_{s/d}=\{1+a_{s/d}^2-\zeta_{s/d}^2+[(1+a_{s/d}^2-\zeta_{s/d}^2)^2+4\zeta_{s/d}^2]^{1/2}\}/2$ 
and $\zeta_{s/d}\equiv\zeta_0/D^{\pm}$. 
Here we have introduced the normalized (average) field amplitude 
$a_{s/d}^2\equiv I_{s/d}/I^*$ with $I^*\equiv m_e^2\omega^2c^3/4\pi e^2$,
the Doppler factors $D^{\pm}=\sqrt{(1\pm\beta)/(1\mp\beta)}\gtreqless 1$, and the surface
density $\zeta_0\equiv\pi n_e\ell/n_c\lambda$, with $n_c\equiv m_e\omega^2/4\pi e^2$
being the critical density. Notice that for a linearly polarized (LP) pulse the peak
intensity $\hat{I}$ is approximately twice the intensity $I$, whereas they coincide 
for a circularly polarized (CP) pulse. 
If both the source and the driver pulse fields are ultrarelativistic ($a_{s/d}\gg1$), 
the reflectivity can be approximated as~\cite{macchiNJP10,macchiPRL09}: 
$\mathcal{R}_{s/d}\approx1$ if $\zeta_{s/d}>a_{s/d}$ and
$\mathcal{R}_{s/d}\approx\zeta_{s/d}^2/a_{s/d}^2$ if $\zeta_{s/d}<a_{s/d}$, 
which presents the reflectivity with accuracy better than 2\% for $a_{s/d}>50$.
Hence, the condition $\zeta_{s/d}>a_{s/d}$ has to be fulfilled 
to secure $\mathcal{R}_{s/d}\approx1$. 

In our model the foil is initially at rest and it is 
accelerated along the positive $x$ direction by the driver pulse. 
In order to determine the value of the Doppler factor after 
the acceleration phase $D^+_0$, we assume that $\zeta_0>a_{d}$ 
and thus $\mathcal{R}_d\approx 1$. The velocity of a foil accelerated 
by the radiation pressure~\cite{mulser-book} 
of the driver pulse can be calculated analytically by 
employing the ``light sail'' equation for 
a perfectly reflecting mirror~\cite{esirkepovPRL04,macchiNJP10,macchiPRL09} 
and the result for $D^+_0$ is $D^+_0=1+\mathcal{E}_d/\zeta_0$, where 
$\mathcal{E}_d=2\pi Zm_e\int{a_d^2(w)\, d w}/Am_p$ 
is the `effective' energy of the driver pulse. 
Here $Z$ ($A$) is the ion atomic number (weight) and $a_d^2(w)=I_d(w)/I^*$ 
is the field amplitude as a function of the foil phase $w=[t/T-x(t)/\lambda]$. 

Since the foil undergoes a recoil due to the radiation pressure of the source pulse, 
the Doppler factor $D^+$ of the foil at the maximum of 
the source pulse intensity is smaller than $D^+_0$. 
On this respect it is convenient to employ a sharp-rising, high-contrast source pulse, 
as those generated with the plasma mirror technique~\cite{thauryNP07,rodelAPB11}. 
By proceeding as for the calculation of $D^+_0$, we obtain
\begin{equation} \label{E1}
D^+ = \frac{D^+_0}{1 + D^+_0\mathcal{E}_s / \zeta_0}=\frac{\zeta_0 (\zeta_0 + \mathcal{E}_d)}{\zeta_0^2 + \mathcal{E}_s (\zeta_0 + \mathcal{E}_d)},
\end{equation}
where $\mathcal{E}_s=2\pi Zm_e\int{a_s^2(w)\, d w}/Am_p$
which, for a sharp-rising pulse, is the part of the source pulse energy before the 
source pulse intensity reaches its maximum (see page~4 for details). 
Since we seek $\mathcal{R}_s\approx 1$, we require $\zeta_s>a_{s}$, 
which provides the constraint $\zeta_0>\zeta_{0,m}$ with 
\begin{equation}  \label{E2}
\zeta_{0,m}=a_{s}[1-\epsilon +\sqrt{(1-\epsilon)(1-\epsilon +4\mathcal{E}_d/a_{s})}]/2 \, ,
\end{equation}
where $\epsilon\equiv\mathcal{E}_s/a_{s}$ accounts for the effect of the recoil. 
In order to maximize the energy and the intensity of 
the reflected pulse at $\mathcal{R}_s\approx 1$ for fixed driver and source pulses, 
we have to maximize the Doppler factor $D^+$ as a function of $\zeta_0$ 
with the condition $\zeta_0>\zeta_{0,m}$. 
From Eq.~(\ref{E1}), $D^+(\zeta_0)$ has a maximum at 
$\zeta^*_0=\mathcal{E}_d\sqrt{\mathcal{E}_s}/(\sqrt{\mathcal{E}_d}-\sqrt{\mathcal{E}_s})$ 
and monotonically decreases for $\zeta_0>\zeta^*_0$. 
Assuming sufficiently small recoil [$\epsilon<1/2$ and $\mathcal{E}_d>\mathcal{E}_s(1-\epsilon)^2/(1-2\epsilon)^2$], 
then $\zeta_{0,m}>\zeta^*_0$ and the maximum $D^+(\zeta_0)$ 
compatible with $\zeta_0>\zeta_{0,m}$ is at $\zeta_{0,m}$, 
and it is $D^+_m=\zeta_{0,m}/a_{s}$. 

Note that, for a flat foil and fixed driver and source pulses, 
both the maximum intensity $I_r=D^{+4}\mathcal{R}_{s}I_s$ 
and energy $E_r\approx I_{r}S\Delta t_s/D^{+2}$ of the reflected pulse are achieved 
at the minimum $\zeta_0$ such that $\zeta_s>a_{s}$, 
i.e. at $\zeta_{0,m}$. Here $S$ is the surface area of 
the focal spot and $\Delta t_s$ is the source pulse duration. 
In fact, for $\zeta_s<a_{s}$ the reflectivity is $\mathcal{R}_s\approx\zeta_s^2/a_{s}^2$ 
thus $I_r=D^{+2}\zeta_0^2I^*$ and $E_r\approx\zeta_0^2I^*S\Delta t_s$, which are 
monotonically increasing functions of $\zeta_0$. 
The fact that there exists an optimal value of the 
surface density has a simple physical interpretation: 
for fixed driver and source pulses, if 
$\zeta_0$ is too large, the foil slows down and the Doppler factor 
is small. If $\zeta_0$ becomes too small, 
the velocity of the foil increases and the reflectivity 
rapidly decreases because $\zeta_s$ tends to vanish. 
Moreover, at $\zeta_{0,m}$ the reflected pulse energy $D_m^{+2}I_{s}S\Delta t_s$ 
is a monotonically increasing function of $I_s$. 
If $\epsilon<1/3$ and $\mathcal{E}_d<a_{s}(1-\epsilon)^2(1-3\epsilon)/4\epsilon^2$, 
i.e. if the effect of the recoil is sufficiently small, 
the maximum reflected pulse intensity $D_m^{+4}I_s$ 
is also a monotonically increasing function of $I_s$. 
For fixed source pulse, the above conditions account for the 
slowdown of the foil due to the recoil, which becomes 
increasingly important for increasing foil velocity [see Eq.~(\ref{E1})]. 
In a three-dimensional (3D) geometry, a paraboloidal 
mirror can focus the source pulse to its diffraction limit. 
Since the laser wavelength is Doppler reduced 
in the rest frame of the foil, 
the reflected pulse can be focused down to $\lambda^2/D^{+2}$ 
and the intensity at the focus is $I_{r,f}=D^{+6}\mathcal{R}_{s}I_{s}S/\lambda^2$. 
If $\zeta_0>\mathcal{E}_d$, $\epsilon<1/4$ and $\mathcal{E}_d<2a_{s}(1-\epsilon)^2(1-4\epsilon)/(1+2\epsilon)^2$, 
the maximum of the intensity at the focus $I_{r,f}$ is achieved at $\zeta_{0,m}$ 
and it is an increasing function $I_s$. In other cases, the 
maximum of $I_{r,f}$ can be a decreasing function of $I_s$ 
or the maximum of $I_{r,f}$ can be achieved at $\mathcal{R}_s<1$. 
However, in these cases a higher intensity \emph{at the focus} 
is achieved at the expense of a lower reflected pulse power 
$P_r=D^{+4}\mathcal{R}_{s}I_{s}S$ and energy 
$E_r\approx D^{+2}\mathcal{R}_{s}I_{s}S\Delta t_s$. 
\begin{figure}
\begin{center}
\includegraphics[width=0.48\textwidth]{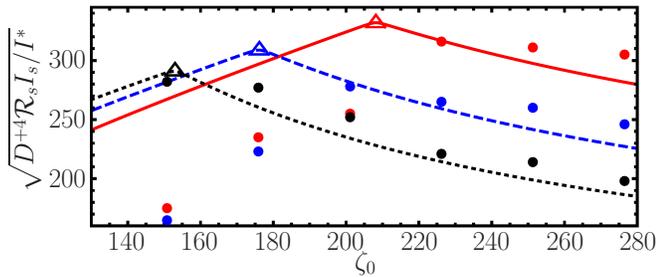}
\caption{(color online). The maximum amplitude of the reflected pulse 
$\sqrt{D^{+4}\mathcal{R}_{s}I_s/I^*}$ as a function of $\zeta_0$
for $a_{d}=130$ and $a_{s}=130$ (solid red line), 
$a_{s}=100$ (dashed blue line) and $a_{s}=80$ (dotted black line).
See the text for further details.} \label{Fig1}
\end{center}
\end{figure}
\begin{figure*}
\begin{center}
\includegraphics[width=0.98\textwidth]{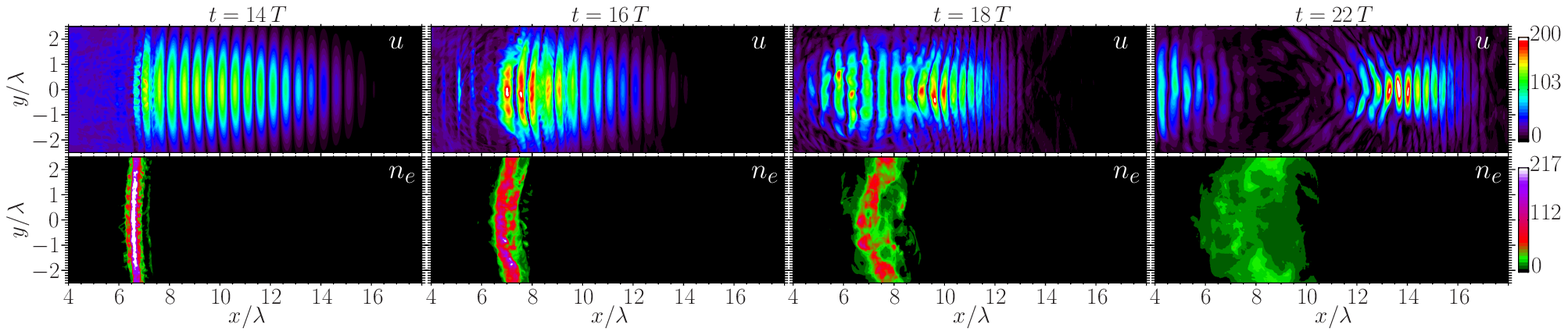}
\caption{(color online). Snapshots of $u=\sqrt{(\mathbf{E}^2+\mathbf{B}^2)/2}$ (first row) 
and $n_e$ (second row) in normalized units, see the text for details.} \label{Fig2}
\end{center}
\end{figure*}

For simplicity, we first consider a driver and source pulse 
with one-cycle $\sin^2$-function rise and fall, 
and with a five-cycle constant plateau. Figure~\ref{Fig1} reports 
the maximum reflected pulse amplitude $\sqrt{D^{+4}\mathcal{R}_{s}I_s/I^*}$ 
as a function of $\zeta_0$ for $a_{d}=130$ and for $a_{s}=130,\,100,\,80$. 
In each case the reflected pulse amplitude initially increases for increasing 
$\zeta_0$, reaches its maximum at $\mathcal{R}_s\approx1$, 
and then decreases as the Doppler factor decreases. 
The three triangles in Fig.~\ref{Fig1} are centered at ($\zeta_{0,m}$, $\sqrt{D_m^{+4}I_s/I^*}$) 
and their position coincides with the maximum of the reflected pulse amplitude, 
confirming our analytical estimates. Since 
in all cases $\mathcal{E}_d<a_{s}(1-\epsilon)^2(1-3\epsilon)/4\epsilon^2$, 
the maximum reflected amplitude rises for increasing $I_s$ (see Fig.~\ref{Fig1}). 
The results of one-dimensional (1D) PIC simulations 
with CP driver and the source pulses are also reported in Fig.~\ref{Fig1} (colored circles), 
the foil being a slab of fully ionized carbon with $n_e=400n_c$. 
The spatial resolution is $\lambda/4000$ and the number of particles per cell per species $1000$. 
Our PIC simulation results agree with the model predictions 
at $\zeta_0>\zeta_{0,m}$, i.e. at $\mathcal{R}_s\approx1$. 

In a multidimensional geometry, the onset of transverse 
Rayleigh-Taylor-like (RT)~\cite{mulser-book} instabilities 
renders the foil `porous' to the source pulse. 
RT instabilities in the radiation pressure acceleration regime 
have been investigated analytically~\cite{pegoraroPRL07,bulanovCRP09}, 
numerically~\cite{pegoraroPRL07,bulanovCRP09,chenPoP11} 
and experimentally~\cite{palmerPRL12}. 
In particular, in Refs.~\cite{pegoraroPRL07,bulanovCRP09} it was shown that 
in the linear approximation the RT instability grows as $\exp[\Phi_d(w)]$ with 
$\Phi_d(w)=\int^{w}_{0}{2\pi[Zm_{e}a_{d}^{2}(u)\lambda/Am_{p}\zeta_{0}\lambda_{RT}]^{1/2}du}$ 
where $\lambda_{RT}$ is the wavelength of the perturbation. 
Our simulations indicate that in order to effectively reflect the source pulse, 
$\Phi_d(w)\lesssim5.7$ for $\lambda_{RT}\approx\lambda$~\cite{palmerPRL12}, 
which can be fulfilled by increasing the value of $\zeta_{0}>\zeta_{0,m}$. 

In our two-dimensional (2D) PIC simulations both the driver and the source pulse have a 
$\sin^2$-function temporal field profile with 15.5~fs duration (FWHM of the intensity),
Gaussian transverse profile and wavelength $\lambda=800\text{ nm}$. 
The driver (source) pulse is CP (LP with the electric field along the $y$ axis) with 
intensity $I_d\approx3.4\times10^{22}\text{ W/cm$^2$}$ ($I_s\approx5.6\times10^{22}\text{ W/cm$^2$}$) 
and spot radius $\sigma_d=3.8\lambda$ ($\sigma_s=1.2\lambda$), 
corresponding to a power $P_d\approx9.9\text{ PW}$ ($P_s\approx1.6\text{ PW}$). 
These parameters are envisaged at the APOLLON laser system~\cite{korzhimanovPU11,dipiazzaRMP12,cheriauxACP12}. 
The foil consists of fully ionized carbon with electron density $n_e=400\,n_c$ and it 
is initially shaped transversely with a thickness distribution 
$\ell=\text{max}[\ell_1,\ell_0\text{exp}(-y^2/2\sigma_f^2)]$, 
with $\ell_1=0.02\lambda$, $\ell_0=0.20\lambda$, $\sigma_f=2.6\lambda$ 
and localized at $x=5\lambda$. Note that the properties of 
such carbon foils can be engineered with high precision nowadays~\cite{krasheninnikovN07,vanDorpME06}. 
It has been shown that Gaussian pulses and shaped foils can be employed 
to generate collimated ion beams~\cite{chenPRL09,chenNJP10}. 
Here we propose to use shaped foils to generate paraboloidal relativistic mirrors. 
Indeed, for $\sigma_d>\sqrt{2}\sigma_f$ the acceleration 
factor $a_d^2(y)/\zeta_0(y)$~\cite{chenPRL09} is larger 
in the outer part of the foil, which therefore takes 
a focusing profile for the source pulse. 
Since for many applications slow focusing and defocusing are desirable, 
we have set $\sigma_d\approx\sqrt{2}\sigma_f$ so the relativistic mirror 
is nearly flat before interacting with the source pulse (see Fig.~\ref{Fig2}). 
The size of the computational box is $20\lambda(x)\times20\lambda(y)$, 
the corresponding grid is $20000(x)\times8000(y)$ and 
900~particles per cell for each species are used. 

\begin{figure}
\begin{center}
\includegraphics[width=0.48\textwidth]{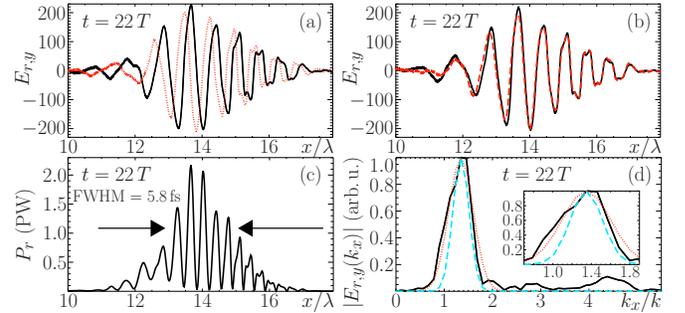}
\caption{(color online). Panel~(a): $E_{r,y}$ along the central axis 
for zero (solid, black line) and $\pi/2$ (dotted, red line) CEP of the source pulse. 
Panel~(b): $E_{r,y}$ with RR effects (solid, black line) and with a preplasma 
on the front surface of the foil (dashed, red line).
Panel~(c): Power contained in a spot with $1\lambda$ radius 
centered on the axis. 
Panel~(d): $|E_{r,y}(k_x)|$ (solid, black line) and the corresponding quantity 
for a Gaussian pulse with two (dotted, red line) and three (dashed, blue line) 
cycles FWHM of the \emph{field} profile. The inset shows a zoom of the main peak region.} \label{Fig3}
\end{center}
\end{figure}

Figure~\ref{Fig2} displays the evolution of the square root of the energy density 
$u=\sqrt{(\mathbf{E}^2+\mathbf{B}^2)/2}$ and of the electron density distribution $n_e$. 
The driver (source) pulse reaches the edge of the foil at $t\approx0$ (10$T$). 
An accurate synchronization between two laser pulses can be achieved, e.g., by generating 
the two pulses from the same seed pulse before the amplification stage. 
Although instabilities have developed ($\Phi_d\approx4.7$ with our parameters) 
and density fluctuations are clearly visible before the source pulse 
impinges on the foil, the foil remains sufficiently compact to 
reflect the first part of the source pulse (see Fig.~\ref{Fig2} at $t\leq16T$ 
and the Supplemental Material~[URL] for a movie of the laser-foil interaction). 
As the source pulse amplitude at the foil position increases, the source pulse `digs through' 
the lower-density regions and abruptly disperses the foil, which becomes 
transparent to the remaining part of the pulse (see Fig.~\ref{Fig2} 
from $t=16T$ to $t=18T$). Finally, at $t=22T$ 
a single few-cycle reflected pulse separated from the foil remnants is observed. 
The peak intensity and peak power of the reflected pulse are: 
$\hat{I}_r\approx2.3\times10^{23}\text{ W/cm$^2$}$ 
(for the source pulse $\hat{I}_s\approx9.6\times10^{22}\text{ W/cm$^2$}$), 
and $\hat{P}_{r}\approx2.2\text{ PW}$, with 5.8~fs duration and 6.8~J energy [see Fig.~\ref{Fig3}(c)]. 
Figure~\ref{Fig3}(a) displays the $y$~component of the electric field of 
the reflected pulse along the central axis for the case of zero (solid, black line) 
and $\pi/2$ (dotted, red line) CEP of the source pulse showing that the reflected pulse 
\emph{inherits} the CEP of the source pulse. 
Inclusion of radiation reaction (RR) effects, according to Refs.~\cite{tamburiniNJP10,tamburiniPRE12}, 
does not significantly alter the reflected pulse [see Fig.~\ref{Fig3}(b)]. 
Our explanation is that when the reflected pulse is generated, the foil density is still high 
and the fields inside the foil are much smaller than in vacuum~\cite{tamburiniNJP10}. 
Moreover, we ensured that the probability of electron-positron pair production remains negligible. 
The influence of a randomly distributed preplasma on the front surface of the foil is also considered 
in Fig.~\ref{Fig3}(b) (dashed, red line). 
The preplasma thickness corresponds to 10\% of the foil thickness and its average density is $n_e/2$. 
The presence of the preplasma reduces the peak intensity, peak power 
and energy of the reflected pulse to $\hat{I}_r\approx1.8\times10^{23}\text{ W/cm$^2$}$, 
$\hat{P}_r\approx2.0\text{ PW}$ and 5.8~J, respectively. 
This can be explained by the increased electron heating due to the enhanced 
penetration of the driver pulse into the preplasma. 
The modulus of the Fourier transform of the $y$~component of 
the electric field along the central axis $|E_{r,y}(k_x)|$, 
where $k_x$ denotes the wave number and $k\equiv2\pi/\lambda$, 
is reported in Fig.~\ref{Fig3}(d) (solid, black line) 
showing that the reflected pulse is chirped and peaked at $\lambda_r\approx593$~nm.
For comparison, the spectra of two Gaussian pulses 
with the same wavelength and with two (dotted, red line) 
and three (dashed, blue line) cycles FWHM of the \emph{field} profile 
are also reported [see Fig.~\ref{Fig3}(d)]. 

In order to account for the slowly-rising profile of the source pulse 
and estimate the wavelength $\lambda_r$ and peak intensity $\hat{I}_r$ 
of the reflected pulse, we approximate the $\sin^2$-function field profile 
with a linearly rising profile $b_{0}w/N$. Here $N$ is the number of cycles 
before the source pulse maximum and $b_{0}=3a_{s}/2\sqrt{2}$ so 
the source pulse and its linear profile approximation have the same 
duration and energy before their maximum. 
Assuming $\mathcal{R}_s\approx1$, the maximum reflected intensity is achieved 
at $\text{min}[N,\hat{w}]$ with $\hat{w}=[4Am_{p}N^2\zeta_0/15\pi Zm_{e}D^+_{0}a_{s}^2]^{1/3}$. 
For a slowly-rising profile $\hat{w}\leq N$, thus $\hat{\mathcal{E}}_s(\hat{w})=\zeta_0/5D^+_{0}$ 
which \emph{does not} depend on the source pulse parameters. Hence, from Eq.~(\ref{E1}) 
we get $\hat{D}^+(\hat{w})=5D^+_{0}/6$. By inserting our numerical parameters we obtain: 
$\lambda_r\approx656\text{ nm}$ and $\hat{I}_r\approx1.2\times10^{23}\text{ W/cm$^{2}$}$ 
for the linearly rising profile, and $\lambda_r\approx593\text{ nm}$ and $\hat{I}_r\approx1.4\times10^{23}\text{ W/cm$^2$}$ 
for the more realistic $\sin^2$-function profile. While $\lambda_r$ is in good agreement with the 2D simulation results, 
$\hat{I}_r$ is underestimated because, by definition, the 1D model does not include focusing effects. 
Indeed, our simulations show that increasing the ratio $\sigma_d/\sigma_f$ by reducing $\sigma_f$ from $2.6\lambda$ 
to $2.4\lambda$ improves the focusing and further enhances $\hat{I}_r$ from
$2.3\times10^{23}\text{ W/cm$^2$}$ to $2.8\times10^{23}\text{ W/cm$^2$}$. 
In addition, higher intensities are expected in a fully 3D geometry 
where, in contrast to 2D simulations, the source pulse is focused also along the $z$~axis. 
We also mention that increasing $P_{s/d}$ by doubling $\sigma^2_{s/d}$ and $\sigma^2_{f}$ 
with the other parameters as reported on page~3 enhances $\hat{P}_r$ to 3~PW
but reduces the intensity enhancement $\hat{I}_r/\hat{I}_s$ from 2.4 to 1.8 because the pulse focusing decreases.

Finally, we stress that even a few-cycle source pulse can be further shortened and amplified. 
Indeed, by employing a $\ell_0=0.17\lambda$, $\sigma_f=2.1\lambda$ shaped foil and 
a driver (source) pulse with 15.5~fs (5.8~fs) duration, $I_d\approx5.1\times10^{22}\text{ W/cm$^2$}$ 
($I_s\approx5\times10^{22}\text{ W/cm$^2$}$) intensity and $\sigma_d=3.1\lambda$ ($\sigma_s=1\lambda$) radius 
[corresponding to a driver (source) power $P_d\approx9.9\text{ PW}$ ($P_s\approx1\text{ PW}$)], 
a single 1.5~cycles (2.1~fs duration), 2~J energy, $\hat{P}_r\approx1.8$~PW and $\hat{I}_r\approx1.4\times10^{23}\text{ W/cm$^2$}$ 
reflected pulse is generated ($\hat{I}_s\approx7\times10^{22}\text{ W/cm$^2$}$). 
Moreover, in contrast to the previous case of a relatively long source pulse, a 2.7~fs duration, 1.3~J energy
1~PW peak power and $4.7\times10^{22}\text{ W/cm$^2$}$ peak intensity transmitted pulse is also generated 
(see the movies in the Supplemental Material [URL]). 
Similar parameters for the driver and source pulses are envisaged at the Extreme Light Infrastructure~\cite{dipiazzaRMP12,eliURL}.

\begin{acknowledgments}
We acknowledge useful discussions with 
B.~M.~Hegelich, N.~Kumar, A.~Macchi and G.~Sarri. 
We thank A.~Macchi for providing his 1D PIC code. 
Some PIC simulations were performed using the computing resources 
granted by the Research Center J\"ulich under the project HRO01. 
\end{acknowledgments}

\bibliography{Few-cycle}

\end{document}